\documentclass[12pt]{article}
\usepackage{graphicx}
 \usepackage{amssymb}
 \usepackage{amsmath}
\usepackage{cite}
 \usepackage{geometry}               
\newcommand {\slsh} [1] {\not{\hbox{\kern-4pt${#1}$}}}

\newcommand {\beq} {\begin{equation}}
\newcommand {\eeq} {\end{equation}}
  \newcommand {\ber}{\begin{eqnarray*}}
  \newcommand {\eer} {\end{eqnarray*}}
\newcommand {\beqn}{\begin{eqnarray}}
  \newcommand {\eeqn} {\end{eqnarray}}

\begin{document}
\begin{titlepage}
\begin{flushright}
{FTPI-MINN-05/41\\
UMN-TH-2413/05\\
}
\end{flushright}
\vskip 0.8cm

\centerline{{\Large \bf Non-Perturbative Yang--Mills}}
\centerline{{\Large \bf
from Supersymmetry and Strings,}}
\vskip 0.4cm
\centerline{\large \bf Or, in the Jungles of Strong Coupling  
\,\footnote{Based on invited 
talks delivered at 
the {\sl  Planck-05/Mohapatra-Fest,} ICTP, 
Trieste, May 23-28, 2005, and {\sl PASCOS-05 
}, Gyeongju, Korea, May 30 - June 4, and  the  
{\sl Cracow School of Theoretical Physics},at 
Zakopane, Poland, June 3-12, 2005.}}
\vskip 1cm
\centerline{\large  M. Shifman  
}
\vskip 0.1cm
\begin{center}
{\em  William I. Fine Theoretical Physics Institute, 
University
of Minnesota, Minneapolis, MN 55455, USA }
\end{center}

\vspace{1.5cm}

\begin{abstract}

\vspace{.3cm}

I summarize some recent developments in the issue of planar equivalence between
supersymmetric Yang-Mills theory and its orbifold/orientifold daughters.
This talk is based on works carried out in collaboration with Adi Armoni, Sasha
Gorsky and Gabriele Veneziano. 
 
 \end{abstract}

\end{titlepage}

\section{Introduction}

Unlike some theories whose relevance to nature is sill a big
question mark, quantum chromodynamics and other
similar strongly coupled gauge theories will stay with us forever. QCD
is a very rich (and quite old) theory supposed to describe the widest range of strong
interaction phenomena: from nuclear physics to Regge behavior at large $E$,
from color confinement to quark--gluon matter at high temperatures/densities
(neutron stars), the vastest horizons of hadronic physics:
chiral dynamics, glueballs, exotics, light and heavy quarkonia and mixtures of thereof,
exclusive and inclusive phenomena, interplay between strong forces and weak interactions, ....
That's why I do not expect theoretical developments to culminate
in full analytic solution of QCD. And yet, in spite of its age,
advances in QCD continue. The most recent advances are due to proliferation
of supersymmetry and string-inspired methods. I will summarize some
recent results which, to my mind, are most promising, and pose some stimulating questions. 

\section{Planar Equivalence}

\begin{figure}[h]
 \centerline{\includegraphics[width=5in]{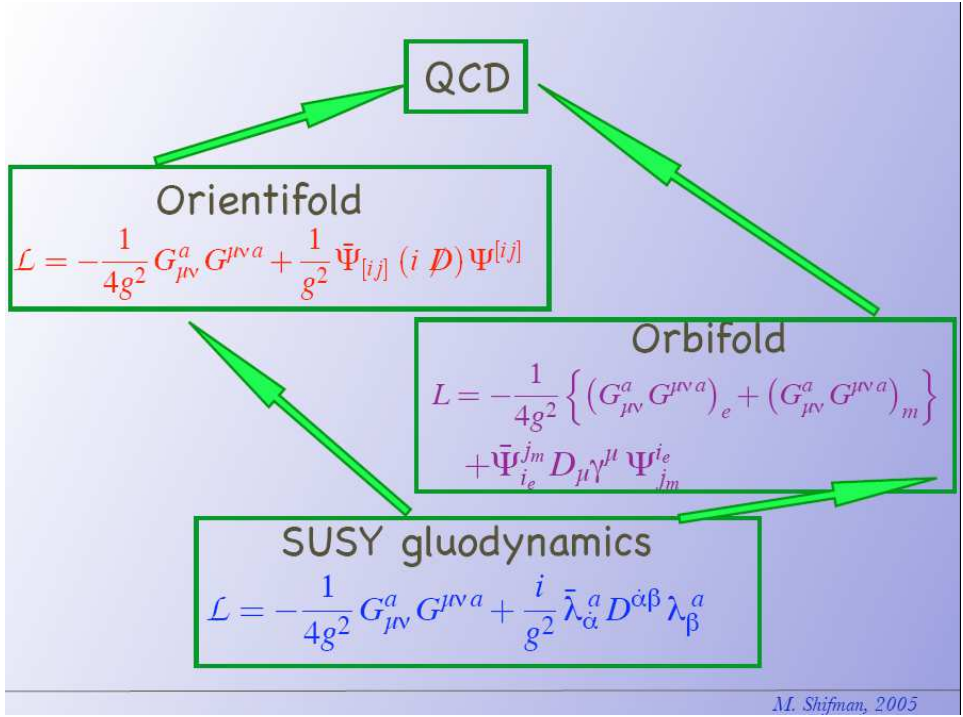}}
 \caption{\small Orbifold and orientifold (orienti-A) daughters are obtained
 from SYM theory by orientifolding and orbifolding, respectively.}
 \label{K1}
 \end{figure}

The main stumbling block in theoretical understanding of strongly
coupled gauge theories is the absence of obvious expansion parameters. 
A hidden parameter which might serve the purpose, $N$ (the number of colors) was suggested by 't Hooft
long ago \cite{thooft}. It governs   expansion in topologies.
The leading order at $N\to\infty$ corresponds to planar
topology. Recently it was realized \cite{strassler,sbuilt} that the very same parameter can be
used to parametrize deviations of certain non-supersymmetric theories,
close relatives of QCD, from supersymmetric Yang--Mills (SYM) theory.
These relatives --- they go under the name orbifold and orientifold
gauge field theories --- are obtained from supersymmetric gluodynamics by means of 
orbifolding and orientifolding, procedures well known in string theory.
For our purposes we do not need to delve in string-theoretic aspects
since all results we need can be readily formulated in field-theoretic language.
They are shown in Fig.~\ref{K1}. SYM theory is assumed to be 
SU($2N$) or SU($N$) gauge 
theory. The first case is pertinent to $Z_2$ orbifolding, the second to orientifolding.
Then the $Z_2$ orbifold daughter\,\footnote{$Z_n$ orbifold daughters with $n>2$ will not be discussed since these theories are chiral and, hence, cannot be considered as close relatives of QCD. Two other lines of research that are marginally
related to my main topic are left aside,
namely (i) orbifold pairs with both theories, parent and daughter, 
supersymmetric; and (ii) orbifolding with one or more 
compact spatial dimensions. In both cases there are special circumstances whose consideration will lead me far astray.} 
has the gauge group 
SU($N$)$\times$SU($N$), and the fermion sector consisting of
one bifundamental Dirac spinor. The gauge coupling of the orbifold daughter is adjusted as follows
\beq
g^2_D = 2 \, g^2_P\,,
\label{231}
\eeq
where the subscripts $D$ and $P$ mark the daughter and parent theories.
For historic reasons the first SU($N$) is often referred to
as ``electric" (and marked by $e$), while the second
as ``magnetic" 
(and marked by $m$).

The orientifold gauge theory is even closer to QCD.
Its gauge group is SU($N$),  the same as in the parent 
SU($N$) SYM theory. The gauge couplings
of the parent SYM and its orientifold daughter are identical too. The fermion sector
consists of one Dirac fermion either in two-index symmetric (orienti-S)
or two-index antisymmetric (orienti-A) representation of SU($N)_{\rm color}$.
In fact, at $N=3$ orienti-A is identical to one-flavor QCD.

Both daughter theories, in the limit $N\to\infty$, were shown \cite{sbuilt}
to be perturbatively equivalent to their parent, supersymmetric gluodynamics. In other words, all planar Feynman graphs of the daughter theories that can be mapped
onto the parent theory are in one-to-one correspondence with
the parent planar graphs.

This remarkable observation motivated 
\cite{strassler} a non-perturbative orbifold (NPO) conjecture, according to which
the above planar equivalence holds beyond perturbation theory, non-perturbatively, 
in a common sector, i.e. the sector of both theories, orbifold and SYM, 
which admits mapping in both directions.

As we will see shortly,
radical distinctions in the vacuum structure of orbifold and SYM theories make NPO planar equivalence unlikely.
At the same time, planar equivalence between orientifold daughter and 
its supersymmetric parent was solidly demonstrated, see
\cite{ASV-one,ASV-two} and the review paper \cite{ASV-three}, with quite a 
few far-reaching consequences that ensued almost immediately. Corresponding results were reported a year ago at various conferences, and I will not discuss them now 
(except for a few marginal remarks), referring the 
interested reader to  \cite{ASV-three}.
Instead, I will dwell on the $Z_2$ orbifold daughter, a theory whose dynamics is rich and interesting irrespective of its (highly probable) non-perturbative non-equivalence
to SYM theory.

\begin{figure}[h]
 \centerline{\includegraphics[width=5in]{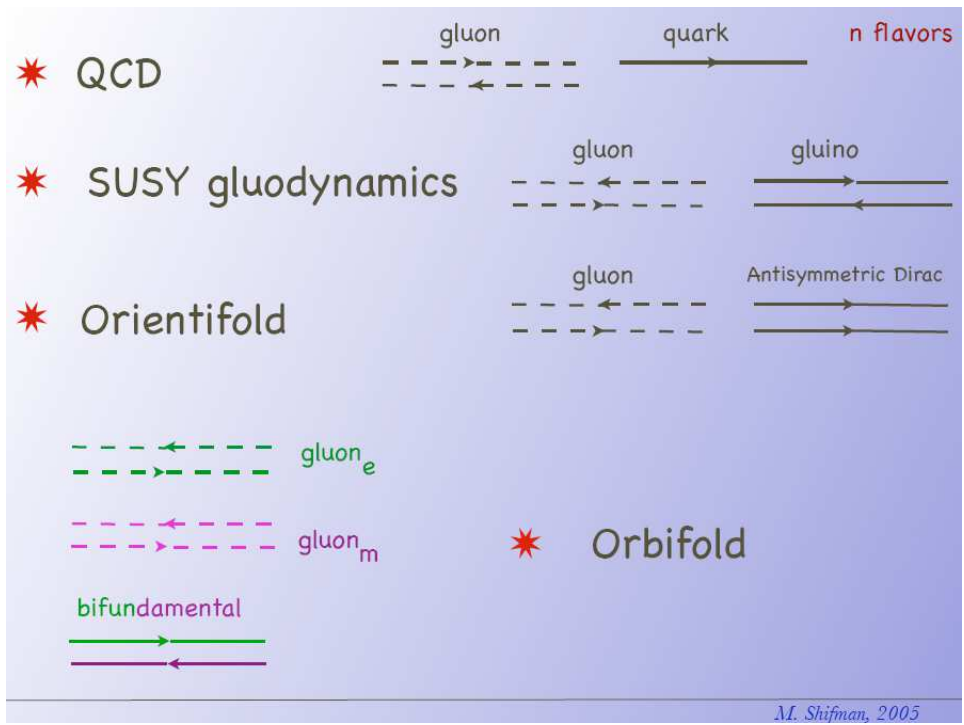}}
 \caption{\small Feynman rules in QCD, SYM theory and orbifold and orientifold (orienti-A) daughters in the 't Hooft notation.}
 \label{K2}
 \end{figure}

Concluding this section I would like to display the 't Hooft large-$N$ diagrammar
for supersymmetric gluodynamics and its daughters
(Fig. \ref{K2}) which makes the perturbative proof of
planar equivalence almost self-evident.

\section{Non-perturbative planar equivalence: what does it mean?}
\label{what}

As we will see shortly, SYM theory and its orbifold daughter have distinct vacuum
structures even at $N =\infty$.
The number of underlying (short-distance) degrees of freedom 
is also different.
Under the circumstances one should carefully define
what is expected to be equivalent. To calibrate the answer to this question it is instructive
to consider an example where the answer is known, namely, let us compare
SYM theory with {\em itself}
in
two cases: 
\beqn
\label{1}
N_c = 2N, \quad g^2_P\equiv g^2\\[2mm]
\label{2}
N_c = N, \quad g^2_D\equiv 2  g^2\,.
\eeqn
The 't Hooft coupling in both cases is one and the same,
$\lambda = 2Ng^2$, which entails, in turn, the equality
of the dynamical scales, $\Lambda_P=\Lambda_D$.
We will refer to the theories (\ref{1}), (\ref{2}) as parent/daughter.  
Having one and the same 't Hooft coupling,
these theories are planar equivalent.
This is as good as it gets, indeed.

Note that for the purpose of
infrared regularization  we will introduce a small gluino mass term
$-(m/g^2)$Tr$\lambda^2 +$ h.c. The mass parameter $m$ is assumed to be real and positive
(a phase can be introduced through the $\theta$ term). The value of $m$
must be the same in both theories since this parameter is physically observable.

Now, both theories are confining and have coincident spectra of composite bosons, 
up to $O(1/N^2)$ corrections. More exactly, we compare here excitation spectra over
vacua which can be mapped one onto another, for instance, those characterized
by real (and negative) gluino condensate $\langle {\rm Tr}\,\lambda^2\rangle$.
One must be careful since the parent theory has $2N$ vacua
while the daughter one $N$ vacua. In fact, the boson spectra in adjacent vacua
differ only by terms $O(m/N^2)$ (one should remember that at $m\neq 0$ 
the true vacuum is the one
in which Tr$\lambda^2$ is real and negative; others are quasistable,
with an exponentially suppressed decay rate \cite{Shifman:1998if},
$\Gamma \sim \exp{(-N^4)}$). 
Only for distant vacua, e.g. those with negative and positive
Re$\langle {\rm Tr}\,\lambda^2\rangle$, the spectra are shifted by $O(N^0)$.
This fact is related to another similar statement. Switching on $\theta\neq 0$
changes bosonic spectra since at $m\neq 0$ particle masses are
$\theta$-dependent.  However, changing $\theta$ from zero to
$\delta\theta\sim 2\pi$ produces an impact on the spectrum suppressed
by $1/N^2$. This can be readily seen e.g. from the SVZ sum rule \cite{SVZ}
analysis
at $\theta\neq 0$,
for instance,
 for highly excited two-gluino bosons one can estimate
\beq
\delta m_n^2 \sim \frac{m\,\Lambda^{-2}}{g^2}\,N_c^{-2}\, {\rm Re}\,
\langle {\rm Tr}\,\lambda^2\rangle \,,
\label{7}
\eeq
where  $n$ is the number of the radial excitation,
$n\gg 1$. CP odd quantities which might be generated at
$\theta\neq 0$ are $O(1/N)$.

The impact of $\theta\neq 0$ becomes of order $O(N^0)$ only if
$\delta\theta\sim 2\pi N$.

At the same time, even though
the bosonic spectra are planar equivalent, the vacuum energy densities are not equal.
 The vacuum energy density\,\footnote{
Tr is normalized in such a way that 
Tr\,$(T^aT^b)=\delta^{ab}$ where $T^a$ are color generators
in the fundamental representation.}
\beq
{\cal E} = m g^{-2} \,\langle {\rm Tr}\,\lambda^2\rangle  \sim N^2  
\label{3}
\eeq 
is sensitive to the overall number of the fundamental degrees of freedom.
It is obvious that we cannot demand the equality of the vacuum energies
in the parent/daughter theories.  Equation (\ref{3}) is fully compatible
with the fact that the $m$ dependence of the  composite boson masses
is identical in the parent and daughter theories.
This can be seen from OPE for the two-point function
\beq
\left \langle  g^{-2} \, {\rm Tr}\, \bar\lambda_{\dot\alpha}\lambda_{\alpha},\,\,
g^{-2} \, {\rm Tr} \, \bar\lambda_{\dot\beta}\lambda_{\beta}\right\rangle
\label{4}
\eeq
which scales as $N^2$. The mass correction to the above two-point function
is given by (\ref{3}). The relative
weight of the leading (unit) operator and the mass correction term is $N$-independent.

Returning to the $\theta$-dependence, the coincidence of the 
parent/daughter boson spectra can be maintained
provided 
\beq
\theta_D = \frac 1 2 \, \theta_P \,.
\label{6}
\eeq
As was mentioned, for $\theta = O(1)$
the vacuum angle effects in the spectra are irrelevant as they are suppressed by
$1/N^2$. However, one can consider $\theta \sim C N 2\pi$
where $C$ is small numerically but not parametrically.\footnote{If $C\sim 1$ one looses the vacuum quasi-stability.} Then Eq.~(\ref{6}) follows from holomorphic dependence
of appropriate quantities on complexified coupling constant
which is dictated, in turn, by supersymmetry of the
parent/daughter theories. I will   further comment on this issue in Sect.~\ref{glance}.

The $\theta$ term has no impact whatsoever on perturbation theory.
It is not seen at all in perturbation theory.
Therefore, perturbative proofs of planar equivalence have nothing to say regarding this aspect. 
On the other hand, non-perturbative quantities, such as the gluino condensate,
do carry a  $\theta$
dependence which imposes the above  identification of the parent/daughter vacuum angles.
The Dashen points
\cite{dashen}  in the parent/daughter theories do not match each other,
as a consequence of a mismatch in the vacuum multiplicities.
It is worth emphasizing that (i) in discussing the $\theta$ evolution
we have to stick, at $\theta \geq\pi$, to a ``wrong" (quasi-stable) vacuum
which will ensure a smooth evolution;
(ii) the Dashen phenomenon is then
irrelevant in the leading in $N$ approximation.

Besides particle excitations both theories have 
domain-wall excitations. The tensions of the elementary walls and their multiplicities
scale as $N$ and are, therefore, different
in the parent/daughter theories. A common factor here is
that all domain walls interpolating between vacua with distinct values
of the gluino condensate are stable. 

\section{A refinement of the proof of planar equivalence for orientifold daughter} 
\label{refinement}

A certain aspect in the previous analysis of non-perturbative planar equivalence between
the SYM parent and orietifold daughter was treated at an intuitive level. 
This gap
is closed in a refined proof  \cite{ASV-two}
making use of the fermion loop expansion.
The equivalence extends to $\theta$ effects, e.g. the topological  susceptibility ---
a feature which is certainly lost in the case of the orbifold daughter.
This is in one-to-one correspondence with the fact that
the vacuum structure of the orientifold
daughter at $N\to\infty$ is identical to that of the parent theory.
In particular, there is an exact matching of the
Dashen transitions.

\section{The role of $Z_2$ invariance in the orbifold theory}

The Lagrangian of the  orbifold theory,
\beqn
{\cal L} &=& -\frac{1}{4g^2}\left[ \left(  G_{\mu\nu}^a G^{\mu\nu\,,a }\right)_e +
\left(  G_{\mu\nu}^a  G^{\mu\nu\,,a }\right)_m
\right]\nonumber\\[3mm]
&+& \frac{1}{g^2}\left[ \overline{\lambda^{i_e}_{j_m}}\left(i\slsh{D}\lambda \right)^{i_e}_{j_m}+
\overline{\lambda^{i_m}_{j_e}}\left(i\slsh{D}\lambda \right)^{i_m}_{j_e}
\right]\,,
\label{261}
\eeqn
has an obvious discrete $Z_2$ symmetry with respect to the interchange
$e\leftrightarrow m$.
(Note that in Fig.~\ref{K1} two Weyl spinors,
$$\lambda^{i_e}_{j_m}\quad {\rm and} \quad  \overline{\lambda^{j_m}_{i_e}}\,,$$
are combined in one Dirac spinor. For a while I will omit the subscript $D$ in the gauge coupling. One should remember, however, that $g_D^2 = 2 g_P^2$, see 
Eq.~(\ref{231}).)

A crucial physical question is whether or not this $Z_2$ symmetry is
spontaneously broken. If it is dynamically broken, the number of vacua is doubled.
As a manifestation of the discrete symmetry breaking, domain walls must emerge,
which interpolate between the vacua related by the broken $Z_2$.
The corresponding order parameters are $Z_2$ odd. For historical reasons, the
$Z_2$ odd sector of the theory is referred to as a ``twisted sector."

If the above $Z_2$ is not broken, the spectrum of the theory in each vacuum can be classified with regards to $Z_2$. For instance, $Z_2$ even particles do not mix with $Z_2$ odd, all domain walls of the unbroken theory are $Z_2$ symmetric, and so on.

The fate of nonperturbative planar equivalence between the 
orbifold theory and its supersymmetric parent is
inseparable from the fate of $Z_2$. As was shown in \cite{Kov1,AGS,Kov2},
if $Z_2$ is unbroken, perturbative planar equivalence 
extends to the nonperturbative level. In the opposite case of the
dynamical $Z_2$ breaking, planar equivalence 
is not expected to survive at the nonperturbative level. 
A shift of the vacuum energy from zero ensues: the vacuum energy density is expected to become negative, see Sects.~\ref{mode}, \ref{domainwall} and \ref{natural}. Other immediately observable consequences refer to
the particle spectrum.
Multiple (parity/spin) degeneracies which would be
inherited from supersymmetric Yang-Mills under planar equivalence,
will be lifted.

To see that this is indeed the case suffice it to note
that if the twisted scalar field\,\footnote{Here and below the normalization of  traces is such that
$$
{\rm Tr}\, G^2 = \sum_{a=1}^{4N^2}G_{\mu\nu}^a\,   G^{\mu\nu\,\,  a}\,,
\qquad {\rm Tr}\,( G^2)_e
= \sum_{a=1}^{N^2}\left( G_{\mu\nu}^a\,   G^{\mu\nu\,\,  a}\right)_e \,,
$$
and so on.}
\beq
T \equiv \left( {\rm Tr} \, G_{ e} ^2 - {\rm Tr} \, G_{  m}^2\right)
\label{iman}
\eeq
develops a $Z_2$-breaking vacuum expectation value, while
its pseudoscalar counterpart 
\beq
\tilde T \equiv \left( {\rm Tr} \, G_{ e}\tilde  G_{ e} - {\rm Tr} \, G_{  m} \tilde G_{  m} \right)
\eeq
 does not, this will be transmitted to the untwisted sector e.g. through a term
 \beq
 \delta {\cal L} = \frac{1}{N} \, \tau^2 \, \sigma\,,
 \eeq
where $\tau$ is a meson for which the interpolating field is
$T$, while $\sigma$ is the dilaton (the corresponding
interpolating operator is $S=    {\rm Tr} \, G_{ e} ^2 + {\rm Tr} \, G_{  m}^2 
$). A vacuum expectation value $\langle\tau\rangle \sim N$
will entail a shift in $\langle\sigma\rangle \sim N$,  which will lead,
in turn, to a shift in the $\sigma$ mass of order $O(N^0)$, not accompanied
by a corresponding shift in the mass of the untwisted
pseudoscalar meson.

Thus, understanding dynamics governing the $Z_2$ symmetry
of the orbifold model is a key to solving the issue
of  nonperturbative planar equivalence in the case at hand. What does today's theory tell us on that?

\section{The mode of $Z_2$ implementation}
\label{mode}

String theory prompts us \cite{klebanov-tseytlin,klebanov} that
in the non-supersymmetric (or ${\cal N}=0$) 
orbifold daughter of ${\cal N}=4$  SYM theory, the $Z_2$ symmetry
is spontaneously
broken above a critical value of the 't Hooft coupling. The orbifold
field theory under consideration can be described by
a brane configuration of type-0 string theory \cite{armoni-kol}.
Type-0 strings contain a closed-string
tachyon mode in the twisted sector. The tachyon couples \cite{klebanov} to
the  twisted field  (\ref{iman})
of the SU$_e(N)\times$SU$_m(N)$ gauge theory. The prediction of 
string theory \cite{klebanov} is that the perturbative vacuum at $\langle T \rangle=0$ is unstable. In the {\em bona fide} vacua a condensate of the form
\beq
\left\langle {\rm Tr} \, G_e ^2 - {\rm Tr} \, G_m ^2 \right\rangle = \pm \Lambda ^4
\label{cotf}
\eeq
must develop.

Of course,  a long way lies between the above string construction and the
orbifold field theory specified in Fig.~\ref{K1} or Eq.~(\ref{261}) {\em per se}.
Therefore, it is natural to address the issue of
the spontaneous $Z_2$ breaking directly in field theory.
In Ref.~\cite{GS} (see also \cite{AGS}, v.1) 
low-energy theorems were suggested as a tool for proving nonequivalence
of the orbifold daughter theories to the parent SYM theory.
These theorems become instrumental
under the assumption of  exact coincidence
between the corresponding vacuum condensates. 
However, as explained in Sect.~\ref{what},
the vacuum condensate coincidence 
is not necessary, generally speaking.
The above-mentioned low-energy theorems
reflect not only the vacuum structure --- they are potentially sensitive to
the number of fundamental degrees of freedom. This aspect was pointed out in \cite{Kov2}. In passing from the orbifold theory to its parent the
number of fundamental degrees of freedom doubles.

Relaxing the requirement of  exact coincidence makes
the low-energy theorems  
uninformative:   allowing for unequal condensates one concludes that these theorems
cannot prove or disprove the $Z_2$ symmetry breaking.

Another argument suggested in \cite{AGS} is based on the domain wall dynamics.
If $Z_2$ was unbroken and NPO conjecture valid,
the domain walls in the orbifold theory that are inherited from
SYM theory would be stable. Apparently, this is not the case.
To discuss the issue in more detail I will have  to briefly review  
what is known of the vacuum structure in the orbifold theory (Sect.~\ref{glance}).

Concluding this section, it is instructive to outline
a possible scenario of the development 
of the tachyonic mode coupled to the twisted operator
(\ref{iman}). Let us give a mass term $m$
to the fermion field in (\ref{261}),
\beq
{\cal L}_m = -m\, g^{-2}\, \bar\Psi\,\Psi\,.
\eeq
This mass term is obviously $Z_2$ invariant.
We will consider $m$ as a free parameter, keeping
the dynamical scale $\Lambda$ fixed. Then, at $m/\Lambda\to\infty$
the fermion field can be integrated out leading to two disconnected
SU($N)$ gauge theories, electric and magnetic.
At finite but large values of $m$, there is a weak connection between
the electric and magnetic theories which can be described by
a (local) operator $m^{-4} \,\mbox{Tr}\,  G^2_e\,\mbox{Tr}\,G^2_m$.
The mass-squared matrix of the electric/magnetic
scalar glueballs takes the form
\beq
{\cal M}^ 2 = \left(
\begin{array}{cc}
\mu_e^2 & \alpha^2\\[1mm]
\alpha^2 & \mu_m^2
\end{array}
\right)
\eeq
where $\mu_e^2 =\mu_m^2 ={\rm const}\, \Lambda^2$
and $\alpha$ is a small parameter proportional to $m^{-2}$. The $Z_2$ invariance of the theory manifests itself in the fact that $\mu_e^2 =\mu_m^2\equiv \mu^2$.
The eigenvalues of ${\cal M}^ 2 $ are $\mu^2\pm \alpha^2$.
The corresponding eigenstates are built of mixtures of the electric and magnetic gluons,
with $Z_2$ parity  $+1$ and $-1$, respectively. 

Now, let us diminish $m$ moving towards $\Lambda$. This enhances interaction between the electric and magnetic sectors, which no longer can be described by a local operator. If at $m=0$ the transition matrix element $\alpha^2$
is larger than the diagonal ones $\mu^2$,   a negative
eigenvalue in the twisted sector emerges. At a certain critical value of
$m$,
$$
m_* \sim \Lambda,
$$
the $Z_2$-odd glueball   becomes massless,
while further decrease of $m$ from $m_*$ to zero
makes the corresponding channel tachyonic causing
condensation of the operator (\ref{iman}) and a radical vacuum restructuring
signifying spontaneous breaking of the $Z_2$ invariance.

One can illustrate the very same statement in a slightly different language of effective
Lagrangians. Indeed, if one approaches the critical value $m_*$ from the large
$m$ side one can describe the vacuum structure by the effective
Lagrangian of the type \cite{MS}
\beq
{\cal L} = S_e\,\ln\frac{S_e}{e} +S_m\,\ln\frac{S_m}{e} +\eta \, S_e\,S_m
\label{msel}
\eeq
where $S_{e,m} ={\rm Tr} \, G_{e,m}^2$, and I put $\Lambda = 1$.
The above Lagrangian is explicitly $Z_2$ invariant.
Of course, it is valid only at $\eta\ll 1$, where the vacuum solution is
$Z_2$ invariant too, $S_e=S_m\approx 1$. Assume that, at a qualitative level,
Eq.~(\ref{msel}) can be extrapolated to $\eta\sim 1$. Then, at $\eta = e$
the vacuum solution is still $Z_2$ symmetric, $S_e=S_m= e^{-1}$, 
but the mass eigenvalue corresponding to $S_e-S_m$ vanishes. Further increase of $\eta$
leads to $Z_2$-asymmetric vacuum solutions while the $Z_2$-symmetric extremum
is no more minimum of the potential.

\section{Vacuum structure of the orbifold daughter at a glance}
\label{glance}


The gauge group of the orbifold theory is a direct product of two SU$(N)$'s.
Correspondingly, it has two vacuum angles conjugated to two distinct non-contractible 
cycles in the space of fields. We will introduce these two vacuum angles as follows:
\beqn
{\cal L_\theta} &=& \frac{\theta_D }{32\pi^2}
\left[ \left(  G_{\mu\nu}^a \tilde G^{\mu\nu\,,a }\right)_e +
\left(  G_{\mu\nu}^a \tilde G^{\mu\nu\,,a }\right)_m
\right]\nonumber\\[3mm]
&+&\frac{\vartheta}{32\pi^2}\left[ \left(  G_{\mu\nu}^a \tilde G^{\mu\nu\,,a }
\right)_e -
\left( G_{\mu\nu}^a \tilde G^{\mu\nu\,,a }\right)_m\right]
\label{tva}
\eeqn
They refer to non-twisted and twisted sectors of the theory,
respectively. Since the parent theory has no twisted sector, the NPO conjecture
requires  $\vartheta =0$. Let us set $\vartheta =0$ for the time being, and focus
on  $\theta_D$.

The order parameter of the parent theory marking its $2N$ vacua, the gluino condensate, is mapped
onto the fermion condensate $\bar\Psi \frac{1}{2}(1-\gamma_5)\Psi$  in 
the daughter theory. Note that this operator is $Z_2$ invariant; 
hence, its nonvanishing vacuum expectation value is insensitive to the spontaneous breaking of $Z_2$.

Following the standard line of reasoning, one can conclude
that the fermion condensate does develop, and has $N$ distinct values,
\beq
\left\langle \bar\Psi \frac{1}{2}(1-\gamma_5)\Psi\right\rangle = {\rm const.}\,
N\,\Lambda^3 \exp\left(i\,\frac{2\pi\, k +\theta_D}{N}\right)\,,\qquad k=1,2,...,N\,.
\label{262}
\eeq
The $N$-valuedness of the fermion condensate is in one-to-one correspondence
with the dependence on $\theta_D/N$ which, in turn, follows from the
consideration of the chiral anomaly. Thus, the fermion condensate marks $N$ distinct chirally asymmetric sectors. 

In the absence of the fermion mass term,
$m=0$, the vacuum angle $\theta_D$ is physically unobservable.
Indeed, at $m=0$ the axial current is classically conserved.
The chiral anomaly then allows one to completely rotate away
the vacuum angle $\theta_D$. No physically measurable quantity can depend on it.
In particular, the vacuum energy is $\theta_D$ independent.
Only if $m\neq 0$, the vacuum angle $\theta_D$ becomes observable.

Note that the dependence of the fermion condensate on $\theta_D$
indicated in Eq.~(\ref{262}) and the relation between the parent and daughter vacuum angles (\ref{6}) are compatible with the $\theta$ dependence of the gluino condensate
in SU($2N$) SYM theory, 
\beq
\langle
\lambda^{a}_{\alpha}\lambda^{a\,,\alpha}
\rangle = -6 (2N)\, \Lambda^3 \exp \left(i\, {\frac{2\pi  k +\theta_P}{2N}}\right)\,,
\,\,\, k = 1,..., \, 2N\,,
\label{gluco}
\eeq
The $g_P^2/g_D^2$ ratio, see Eq.~(\ref{231}),  also matches. Thus, the fermion condensate of the orbifold theory could have been projected from the parent theory
provided that the NPO conjecture was valid.

\begin{figure}[ht]
\centerline{\includegraphics[width=3.5in]{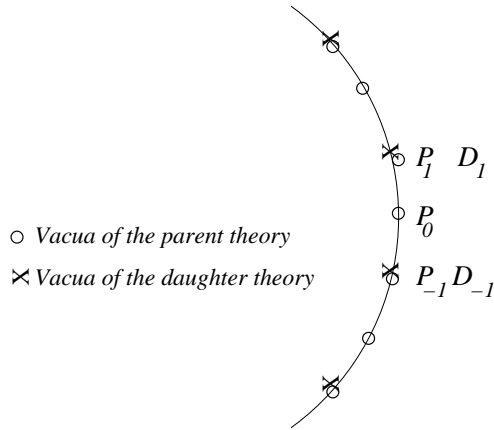}}
\caption{\footnotesize
The vacuum structure in the SU($2N$) SYM theory and its SU$(N)\times$SU($N$) orbifold daughter. Shown is the complex plane of the order parameters,
the gluino condensate $-\langle
\lambda^{a}_{\alpha}\lambda^{a\,,\alpha}
\rangle$ and  the fermion condensate $\, -\left\langle
\bar\Psi\frac{1}{2}\left( 1-\gamma_5\right)\Psi \right\rangle$, respectively.}
\label{ORBtwo}
\end{figure}

As an example,
I depicted the chiral condensates of the parent/daughter theories
at 
\beq
\theta_D = \pi\,,\qquad \theta_P= 2\pi\,,
\label{ili}
\eeq
in Fig.~\ref{ORBtwo}. $P_{0,\pm 1}$
are the vacua of the SYM theory, while $D_{\pm 1}$ are the vacua of the orbifold theory.
Since the vacua reflect the discrete chiral symmetry breaking,
$2N$ vacua of SYM theory are degenerate, and so are $N$ vacua
of the orbifold theory. Whether or not they are degenerate between themselves,
depends on the validity of planar equivalence.

Introducing $m\neq 0$ one  lifts the vacuum degeneracy. For instance, for real and positive 
$m$ the  vacua  $P_{\pm 1}$ are excited (quasistable) because
$$
{\cal E}_{P_{\pm 1}} > {\cal E}_{P_{0}}\,.
$$
At the same time, the daughter theory has two-fold degeneracy, 
$$
{\cal E}_{D_{+ 1}} ={\cal E}_{D_{- 1}},
$$
a phenomenon well-known
at $\theta =\pi$, the so-called Dashen
phenomenon \cite{dashen}. As was explained in Sect.~\ref{what},
$\theta$-dependent effects are suppressed by $1/N$.

\begin{figure}[ht]
\centerline{\includegraphics[width=3in]{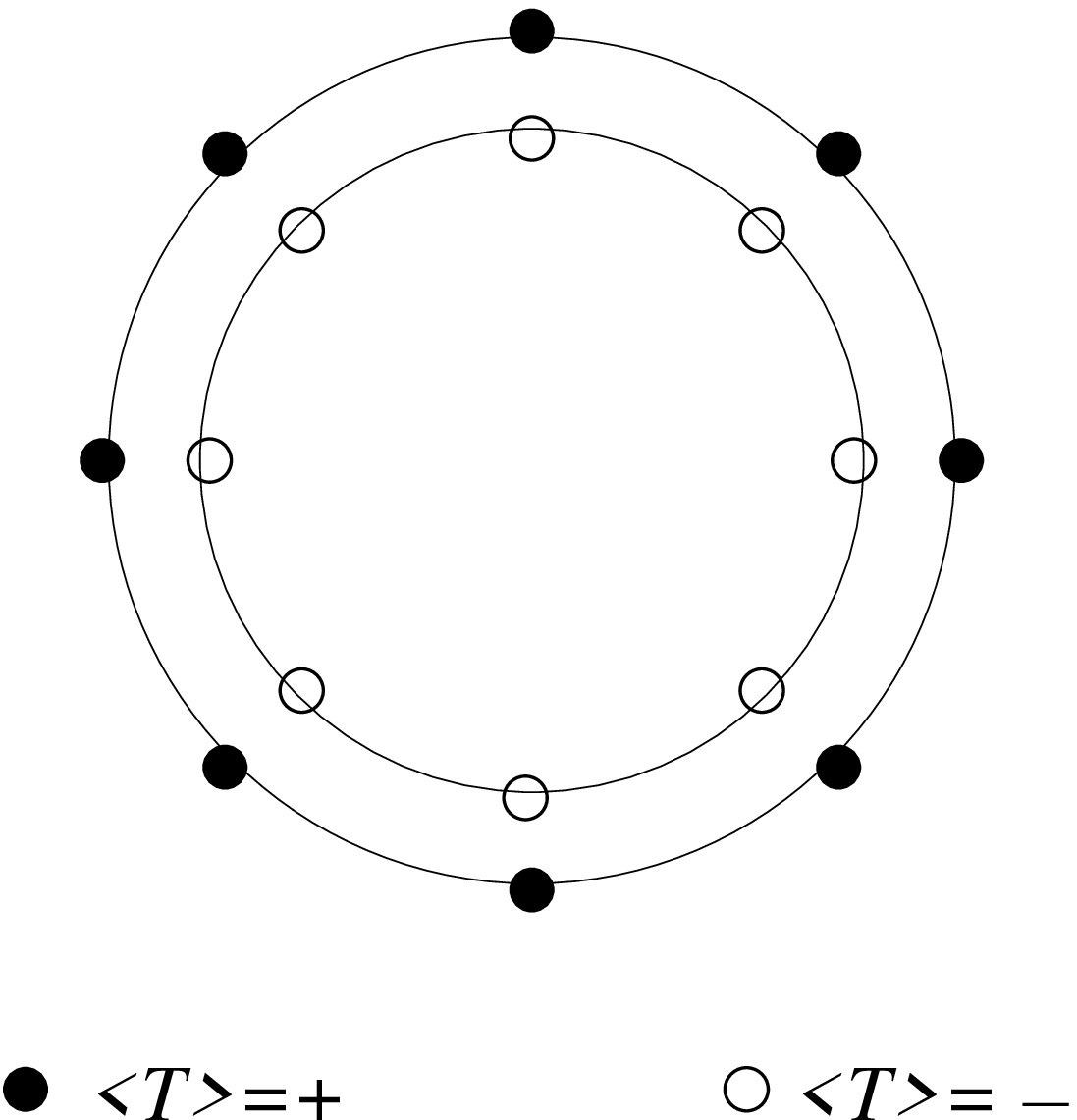}}
\caption{\footnotesize The vacuum structure of the SU(8)$\times$SU(8) orbifold theory reflecting spontaneous breaking of the $Z_2$ symmetry. } 
\label{biman}
\end{figure}

The fermion condensate (\ref{262}) is a good order parameter
for the chiral symmetry breaking. It cannot serve,
however, as an order parameter for the $Z_2$ breaking. 
In the orbifold theory with the spontaneously broken $Z_2$
the fermion condensate does not
differentiate those vacua which are connected to each other
by $Z_2$ because 
it is $Z_2$-even. We must supplement (\ref{262})  by a $Z_2$-odd expectation value
of (\ref{iman}). This vacuum expectation value (VEV) is dichotomic. 
The fermion condensate (\ref{262}) in conjunction with
$\langle T\rangle =\pm \Lambda^4$ fully identifies each  degenerate vacuum
of the orbifold theory. 
If spontaneous breaking of the discrete chiral symmetry 
produces $N$ vacua, this number is doubled
in the process of $Z_2$ breaking.
Somewhat symbolically, the corresponding vacuum structure
is presented in Fig.~\ref{biman}. The angular 
coordinate represents the phase of  (\ref{262}), while the radial coordinate
can take two distinct values
representing the  dichotomic parameter $\langle T\rangle$.

The orbifold theory has a remarkable feature:
because of its proven {\em perturbative} planar equivalence to SYM theory,
the vacuum energy density --- certainly a $Z_2$ symmetric
parameter --- can and does play the role of the order parameter for $Z_2$ breaking.
The vacuum energy density is proportional to the
vacuum expectation value of the 
$Z_2$-even gluon operator
$\langle {\rm Tr} \, G_e^2 +{\rm Tr} \, G_m^2\rangle$. Indeed, this operator
is related, in turn, to the
total energy-momentum tensor of the theory,
\beq
\theta_\mu^\mu =  - \frac{3N }{32\pi^2}
\sum_{\ell = e,m}\left( G_{\mu\nu}^a\, G_{\mu\nu}^a\right)_\ell\,.
\label{beiman}
\eeq
Ever since the gluon condensate was introduced in non-Abelian gauge 
theories \cite{SVZ} people tried
to identify it as an order parameter. Nobody succeeded. The orbifold theory is the
one where it is an order parameter, albeit in a special sense.

If $Z_2$ is unbroken, the orbifold theory is perfectly equivalent at $N\to\infty$
to SYM theory, and then
$\langle {\rm Tr} \, G_e^2 +{\rm Tr} \, G_m^2\rangle$ reduces to
$ \langle {\rm Tr} \, G^2_{\rm SYM}\rangle$.
The latter condensate vanishes due to  supersymmetry of the parent theory.
Hence, the $Z_2$ symmetric vacua in the
daughter theory would have vanishing vacuum energy 
density in the leading order in $N$.

When the $Z_2$-symmetric point becomes unstable,
the  $Z_2$-asymmetric vacua must have a negative energy density.
Equation (\ref{beiman}) implies then that in the genuine vacua
\beq
\langle {\rm Tr} \, G_e^2 +{\rm Tr} \, G_m^2\rangle > 0
\label{gdesim}
\eeq
at order $O(N^2)$.
In this way the gluon condensate acquired the 
role of a $Z_2$ breaking order parameter, much
in the same way as $ \langle {\rm Tr} \, G^2_{\rm SYM}\rangle$
is the order parameter for supersymmetry breaking in SYM theory.

\section{Domain-wall-based argument for $Z_2$ breaking}
\label{domainwall}

In this section we analyze  the domain wall dynamics  in the $Z_2$
orbifold theory.  Since domain walls are ``QCD D-branes" \cite{witten1997}
a similarity between the wall dynamics and D-brane dynamics is clear.

Why domain walls? As well-known,   domain walls
are physical manifestations of spontaneously broken 
discrete symmetries.
Since our consideration aims at exploring the
$Z_2$ breaking in the orbifold daughter theory, addressing  domain
walls is an adequate maneuver. 

To begin with, let me recapitulate
the domain wall topic in the parent theory. SU$(2N)$
SYM theory has BPS domain 
walls \cite{dvali-shifman} that carry both tension $\sigma$ and charge $Q$ (per unit area), with $\sigma = Q$. The expressions for the tension and charge 
can be written as follows \cite{armoni-shifman}:
\beqn
&&
\sigma = {3(2N) \over 32 \pi ^2}\, \int_{\rm wall} \, dz\, {\rm Tr}\, G^2\,, \label{sigmas}\\[2mm]
&&
Q= {3(2N) \over 32 \pi ^2}\,  \int_{\rm wall} \, dz\, {\rm Tr}\, G\tilde G\,,
\label{sigmaq}
\eeqn
where $z$ is the direction perpendicular
to the wall plane. Equation (\ref{sigmas}) is a consequence of the scale anomaly.
The walls interpolating between the adjacent vacua
(e.g. $P_0$ and $P_1$ in Fig.~\ref{ORBtwo}) are called elementary, or 1-walls.
One can consider  bound states of  the elementary walls too. These walls
interpolate between the vacua $i$ and $i+k$ with $k>1$ and are referred 
to as $k$-walls.  
For instance, the wall  interpolating between $P_{-1}$ and $P_1$ in Fig.~\ref{ORBtwo}
is a 2-wall. At $N=\infty$ it is marginally stable,
since the tension of the 2-wall is twice the tension of the 1-wall.
Although elementary walls do interact via the exchange of glueballs, there is an exact cancellation between the contribution of even- and odd-parity glueballs
\cite{armoni-shifman} at $N=\infty$.  From the world-sheet
theory standpoint,  the no-force result is due to the Bose-Fermi degeneracy 
on the wall. I will return to the world-sheet theory
shortly, after a brief remark 
regarding generalizations of Eqs. (\ref{sigmas}) and (\ref{sigmaq})
in the orbifold daughter theory.

For the
tension and charge of the orbifold theory
domain walls one can write\,\footnote{In SYM theory such integrals
are well-defined since $\langle G^2\rangle$ vanishes in any supersymmetric
vacuum.  In the orbifold theory this is not necessarily the
case. The integrals in Eq.~(\ref{tqdt})
must be properly regularized.}
\beqn
\sigma _D &=& {3N \over 32 \pi ^2 }\,  \int_{\rm wall} \,  dz \, {\rm Tr} \,  G_e^2 + {3N \over 32\pi ^2 }\,
\int_{\rm wall} \, dz\,  {\rm Tr}\,G_m ^2 \,,
\label{tqdt} \\[2mm]
Q _D  &=& {3N \over 32\pi ^2 }\,  \int_{\rm wall} \, dz \, {\rm Tr}\left(  G\tilde G\right)_e + {3N \over 32\pi ^2 }\,  \int_{\rm wall} \, dz \, {\rm Tr}\left(  G \tilde G\right)_m\,.
\label{tqdtp}
\eeqn
It is suggestive to think of the domain walls in the orbifold
theory as of marginally bound states of  fractional ``electric" and ``magnetic" domain walls,  with the following tensions and charges:
\beqn
\sigma_e &=&  {3N \over 32 \pi ^2 }\,  \int \,  dz \, {\rm Tr} \, G_e^2 \,,  \qquad  \sigma _m =  {3N \over 32 \pi ^2 }\,  \int \,  dz \, {\rm Tr} \, G_m ^2 \,,
\nonumber \\[2mm]
Q_e &=&  {3N \over 32 \pi ^2  }\,  \int \,  dz \, {\rm Tr} \, (G\tilde G)_e \,,
\qquad Q _m =  {3N \over 32 \pi ^2  }\,  \int \,  dz \, {\rm Tr} \,  (G \tilde G)_m\,.
\label{tqdtp2}
\eeqn
Assuming unbroken
 $Z_2$ s i.e.
$
\sigma _e = \sigma _m\,,
$
we would get
\beq
\sigma_{e,m} = {1\over 2 }\left(\sigma_{e}+\sigma_{m}
\right) ,
\eeq
i.e., a  fractional amount of tension, in full analogy with   fractional D-branes.
Then we would have to conclude that, say, at
$k=2$ two parallel  electric domain walls do not interact at  $N=\infty$.
The same would be valid for the
magnetic walls. Unfortunately, the world-sheet theory
in the case at hand  does not support
this conclusion.

I will again start from SU($2N$) SYM theory.
The world-sheet theory for $k$-walls in ${\cal N}=1$ gluodynamics was derived 
by Acharya and Vafa \cite{AV}.  It was shown to be a (2+1)-dimensional $U(k)$  theory with  level-$2N$ Chern--Simons term and it was shown to have (2+1)-dimensional 
${\cal N}=1$
supersymmetry. 
The action of the theory is
\beq
S=\int d^3 x \left\{ {\rm Tr} \left (
-{1\over 4e^2} F^2 + {2N\over 16\pi} \epsilon ^{ijk} A^i F^{jk}+
{1\over 2} (D_i \Phi)^2 \right ) + {\rm fermions} \right\}\,.
\eeq
All  fields in the action, including the fermion fields, transform in the adjoint
representation of $U(k)$. For definiteness, we will consider a minimal case $k=2$.

In the orbifold daughter,  the
world-sheet theory becomes, by virtue of
the orbifold procedure, a U$_e(1)\times$U$_m(1)$ gauge theory with a neutral scalar
field  and  ``bifundamental"
fermions. The same conclusion on the world-sheet  theory can be obtained 
directly through a consideration of type-0 string theory similar to that
of Acharya and Vafa. In this case the world-sheet action is
\beqn
S  &=&
\int d^3 x \left\{ \sum_{\ell=e,m}\left(  -{1\over 4e^2} F_\ell  ^2 + {N\over 16\pi} \epsilon ^{ijk} A_\ell^i \,
F_\ell ^{jk}+  {1\over 2} (\partial_i\Phi_\ell)^2 \right)\right.
\nonumber\\[3mm]
   &+&\left.  \bar \Psi   \left(\Phi_e - \Phi_m\right) \Psi + ...  \right\}.
\label{WVorbifoldp}
\eeqn
The occurrence of the Yukawa coupling
$
\bar \Psi   \left(\Phi_e - \Phi_m\right) \Psi \label{yc}
$
in the daughter theory, with no counterpart
in the parent one,  is the  fact of a special importance.

One can  interpret  the above expression as follows.
The daughter wall is a sum of the electric  and  magnetic  walls that interact with each other
via the bifundamental fermions. The  electric  branes can be separated from the  magnetic  branes as is seen  from the fact that the Yukawa term $
\bar \Psi   \left(\Phi_e - \Phi_m\right) \Psi \label{yc}
$
 in the action \eqref{WVorbifoldp} can make
the bifundamental fermion massive. The vacuum expectation values 
\beq
\langle \Phi_e \rangle = v_e\,, \qquad \langle \Phi_m\rangle =v_m\, ,
\eeq
which can be chosen to be real
are in one-to-one correspondence with the wall separation. If $v_e \neq v_m$
a mass $\mu$ for the world-sheet fermions is generated,
\beq
\mu  = v_e - v_m\,.
\label{mfwvf}
\eeq
At $\mu \to \infty$ the fermions decouple --- we have two
decoupled U(1) theories.  The world-sheet theory on the separated  electric 
(or  magnetic) domain walls
is just a bosonic  U(1)  gauge theory with a level-$N$ Chern--Simons term.
It is not supersymmetric. There is no reason for the wall tension non-renormalization
and the no-force statement.

The above  conclusion can be backed up by a 
calculation of the wall repulsion \cite{AGS,Zar}.
Needless to say, this repulsion is in contradiction with the NPO conjecture.

\section{Back to the bulk theory}
\label{back}

\begin{figure}[ht]
\centerline{\includegraphics[width=1.5in]{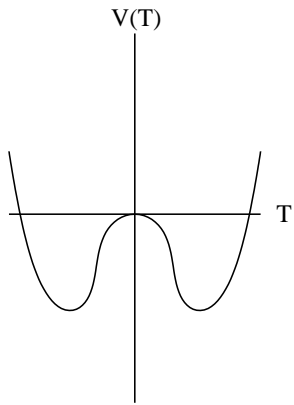}}
\caption{\footnotesize The tachyon field potential. } 
\label{potential}
\end{figure}

If the orbifold has $Z_2$-odd vacua,
the tachyon field potential must have  minima away from the origin,
as shown in Fig.~\ref{potential}, cf. the last paragraph in Sect.~\ref{glance}.
String theory gives us a hint that the point $T=0$ is unstable.
Field theory allows us to say that the potential
 $V(T)$ is bounded from below since  the regime of large expectation values is fully controlled by semiclassical dynamics. 
 
From the field-theoretic standpoint
it is clear that the only possibility open is that in the {\em bona fide} vacuum $\langle T\rangle\sim \Lambda ^4$.
Non-stabilization of tachyons would mean $\langle T\rangle\gg \Lambda ^4$, which is ruled out.

In the parent  SYM theory with the gauge group
SU$(2N)$, there are $2N$ vacua, with the gaugino condensate as an order parameter,
see Fig.~\ref{ORBtwo}. The domain walls interpolate between these $2N$
various vacua. In the daughter theory the situation is more complicated. $Z_2$ breaking implies that
each vacuum of the $N$ ``false" perturbative vacua splits into two,  see 
Fig.~\ref{biman}. 

A scenario of 
the wall inheritance from the parent to   daughter theory we have in mind is as
follows. We first pretend that the daughter theory is planar equivalent to SYM, and that the $Z_2$ symmetry is unbroken.  Start from a
2-wall in the parent theory. It will be inherited, as  a minimal
wall in the daughter theory.
This is seen from Fig.~\ref{ORBtwo}. 
We may consider e.g. the wall connecting $D_{-1}$ and $D_{1}$ in the daughter (this is a minimal wall in the daughter),
versus the wall connecting $P_{-1}$ and $P_{1}$ in the parent (this is a 2-wall in SYM
theory).

In the parent theory two 1-walls comprising the 2-wall
do not interact with each other (at $N=\infty$).
If we consider them on top of each other, the world-volume
theory has $U(2)$ gauge symmetry.
However, nobody precludes us from introducing a separation.
Then we will have U(1) on each 1-wall, U(1)$\times$U(1) altogether. The tension of each 1-wall is 1/2
of the tension of the 2-wall, it is well-defined and receives
no quantum corrections.
The fact that the world-volume
theory on each 1-wall is supersymmetric is
in one-to-one correspondence with the absence of quantum corrections.
In the daughter theory the minimal wall splits into one
electric and one magnetic repelling each other.
(The electric one connects $D_{-1}$ with the would-be vacuum which is a counter-partner of $P_0$, the magnetic one connects the
would-be vacuum which is a counter-partner of
$P_0$ with $D_{1}$). 

How can one visualize this situation?

In the parent theory we have degenerate minima at all points
$P_i$. In the $Z_2$ broken orbifold theory these minima become saddle point
(still critical points, but unstable). Near every second saddle point two minima develop.
Of course, the walls that would be inherited from SYM are all unstable,
with tachyonic modes. 1-walls are transformed into electric/magnetic
walls of the orbifold theory, which are still unstable and, in fact, decay.
Each of them separately could decay only into a ``twisted wall"
connecting white and adjacent black true vacua.
The ``untwisted'' electric+magnetic wall can decay into a minimal stable wall of the daughter theory which connects two neighboring black vacua or
two neighboring white vacua.

\section{Why non-perturbative non-equivalence is natural?}
\label{natural}

I this section I will try to illustrate
why a shift of the vacuum energy from zero is expected in the
orbifold theory. Needless to say this can only happen if perturbative planar equivalence
gives place to 
non-equivalence at the non-perturbative level.
The issue to be discussed here is the vacuum angle dependence, see Eq.~(\ref{tva}).
In this section I will treat $N$ as a fixed parameter
assuming that transition to $N\to\infty$ is smooth, as is the case in
pure Yang-Mills theory.

As was mentioned, physical quantities do not depend on $\theta_D$,
as this angle can be rotated away. A weak dependence appears if $m\neq 0$,
but we will be interested in the limit $m\to 0$.
For our present purposes   $\theta_D$ is irrelevant and can be set at zero.

Unlike $\theta_D$, the second vacuum angle, $\vartheta$,
cannot be rotated away: the only axial current of the theory is
$Z_2$ even while the $\vartheta$ term in Eq.~(\ref{tva}) is $Z_2$ odd.
Thus, physics must be $\vartheta$ dependent even at $m\to 0$.
Of course, at the   end of the day we want to focus on the  $\vartheta =0$ sector. 
Nothing precludes us, however, from dealing with
$\vartheta \neq 0$ sectors at intermediate stages of our consideration.
Knowledge of pure Yang--Mills theory and Yang-Mills theory with massless quarks
can be used as a reference frame and a guiding principle.

In  pure Yang--Mills theory the vacuum angle reflects a non-trivial topology in the space of fields and the possibility of tunneling \cite{oldinst}, a nonperturbative effect which makes the vacuum energy $\theta$ dependent and decreases the vacuum energy at
$\theta =0$. Instantons  exemplify the tunneling trajectories \cite{SPo}.
Massless quarks suppress instantons (and any other field configurations
with nonvanishing topological charges), freeze tunneling and make physics 
(including the vacuum energy) $\theta$ independent. Likewise, in
SYM theory instanton does not contribute to the vacuum energy
because of the gluino zero modes (an instanton-antiinstanton configuration
could contribute but it has a vanishing topological charge and 
is topologically unstable.)

\begin{figure}[ht]
\centerline{\includegraphics[width=2.5in]{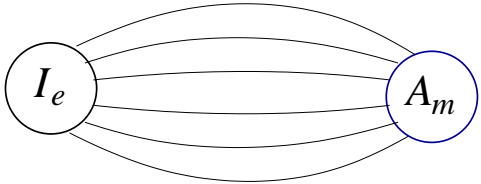}}
\caption{\footnotesize Topologically stable instanton-antiinstanton
pairs in the orbifold theory. Instanton belongs to the electric SU($N$)
while antiinstanton to the magnetic SU($N$).} \label{bbiman}
\end{figure}

In the orbifold theory we have two topological charges.
Massless bifundamental fermions do suppress tunneling in the direction conjugate
to $\theta_D$. That's why  physics cannot depend on this parameter.
However, the orbifold theory exhibits a new phenomenon: topologically  stable
 instanton-antiinstanton
pairs, connected through fermion zero modes, see Fig.~\ref{bbiman}.
The stability is due to the fact that they belong to distinct gauge factors.
Therefore, although the overall topological charge (electric + magnetic)
vanishes
(all fermion zero modes are contracted), still
instanton$_e$ cannot annihilate antiinstanton$_m$.
The ``twisted" topological charge, conjugate to $\vartheta$,
is the difference between the electric topological charge and the magnetic one.
Non-trivial topology and tunneling with regards to the twisted 
topological charge is not suppressed by massless fermions.

That's why  physics does depend on $\vartheta$.
With regards to $\vartheta$ effects, the orbifold theory is expected to be 
similar to pure Yang--Mills, with no
massless  quarks. The instanton$_e$-antiinstanton$_m$ pair plays the
role of the instanton in pure Yang--Mills.
In particular, the vacuum energy ${\cal E}$ becomes a function of $\vartheta$ (more exactly, $\vartheta /N$), and, if so,
there is absolutely no reason 
for  ${\cal E}(\vartheta ) =0 $ at $\vartheta =0$.

In fact, one is expected to find ``vacuum families,"
of the type described by Witten \cite{W} (see also \cite{Shifman:1998if}):
a group of $\sim N$ quasistable ``vacua" entangled in the process of
$\vartheta$ evolution and interchanging their position
each time $\vartheta$ reaches $k\pi$ where $k$ is 
integer.\footnote{This is in addition to $N$ chiral sectors
labeled by $\langle\bar\Psi \frac{1}{2}(1-\gamma_5)\Psi\rangle$.
Note that the first crossover Dashen point is at $\vartheta =\pi/2$.}
The issue of the dynamical $Z_2$ breaking in this language is formulated as follows:
at $\vartheta =0$ each vacuum family contains two degenerate stable vacua
connected by a $Z_2$ transformation. At generic $\vartheta \neq 0$
the  $Z_2$ symmetry of the action is explicitly broken by the 
$\vartheta$ term in Eq.~(\ref{tva}).

\section{Conclusions}
\label{conclusions}

Examples of cross-fertilization between string theories and gauge field theories are abundant. The topic of planar equivalence 
between supersymmetric and non-super\-symmetric gauge theories
emerged in this way.
In the recent years it produced quite a few spectacular results and stimulated
various activities in diverse directions. Two classes of 
non-supersymmetric models were identified as daughter theories:
orbifold and orientifold. Planar equivalence is valid for both
at the perturbative level.

In this talk I tried to summarize
recent  nonperturbative analyses of the orbifold theories.
It was found, beyond reasonable doubt,
that the $Z_2$ symmetry of the  $Z_2$ orbifolds
is the key to nonperturbative planar equivalence.
If it is not dynamically broken,  planar equivalence
must extend to the nonperturbative level.
The opposite is also true: spontaneous breaking of 
$Z_2$ entails a nonvanishing vacuum energy and a failure of 
 planar equivalence. I discussed arguments in favor of 
nonperturbative nonequivalence such as  domain wall dynamics and 
$\vartheta$ dependence.
Unfortunately, there is no iron-clad proof of the statement.
At a certain point, low-energy theorems seemed to provide such a
proof. It turned out, however, that they may or may not be relevant
since they are sensitive not only to the vacuum structure
of the parent/daughter theories, but also to the number of the
fundamental degrees of freedom which is different in the parent/daughter theories.

In this sense,  situation  with the orientifold daughter theories
is much more favorable. Nonperturbative  planar equivalence certainly
does hold for the orientifold theories. Why they are better than their
orbifold cousins?

String theorists are familiar with this phenomenon. Type-II strings on orbifold singularities of the form $\,C^3/ Z_n\, $, or type-0 strings   always 
contain a tachyon in the twisted sector (and fractional branes).

For orientifold theories the situation is conceptually different. This
nonsupersymmetric gauge
theory  has  no twisted sector and, in particular, it does not contain fractional domain walls; hence, it is guaranteed that the
theory inherits its vacua from the SUSY parent.

Similarly, the candidate for a string dual of the orientifold theory ---
Sagnotti's type-$0^\prime$ model \cite{sagnotti} --- contains 
no tachyon since it was projected out by  orientifolding.
 
The orientifold theory is closer to QCD. On the other hand, the
orbifold theory has rich internal dynamics presenting, in a sense,  a hybrid between
QCD with massless quarks and pure Yang--Mills. Even though its planar equivalence to SYM theory is highly unlikely, it is an alluring   target for future studies. 

\section*{Acknowledgments}

I am grateful to Adi Armoni and Sasha Gorsky
for valuable comments on the manuscript.
This work was supported in part by DOE grant DE-FG02-94ER408.

\end{document}